\documentclass[aps,prl,amsmath,amssymb,twocolumn,nofootinbib, showpacs, showkeys, nofootinbib]{revtex4}
\usepackage{pslatex,graphicx,dcolumn,bm,natbib}
\usepackage{multirow}
\usepackage{subfigure}
\date{\today}

\begin{document}

\title{Spatiotemporal Chaotic unjamming and jamming in granular avalanches}

\author{Ziwei Wang$^{1}$ and Jie Zhang$^{2,3}$}
\email{jiezhang2012@sjtu.edu.cn}
\affiliation{$^1$Zhiyuan College \& $^2$Institute of Natural Sciences \& $^3$Department of Physics and Astronomy, Shanghai Jiao Tong University, Shanghai 200240, China
}

\begin{abstract}
 We have investigated the spatiotemporal chaotic dynamics of unjamming and jamming of particles in a model experiment -- a rotating drum partially filled with bidisperse disks to create avalanches. The magnitudes of the first Lyapunov vector $\delta u(t)$ and velocity $v(t)$ of particles are directly measured for the first time to yield insights into their spatial correlation $C_{\delta u,v}$, which is on statistical average slightly larger near the unjamming than the value near the jamming transition. These results are consistent with the recent work of Banigan et al\cite{banigan2013chaotic}, and it is for the first time to validate their theoretical models in a real scenario. $v(t)$ shows rich dynamics: it grows exponentially for unstable particles and keeps increasing despite stochastic interactions; after the maximum, it decays with large fluctuations. Hence the spatiotemporal chaotic dynamics of avalanche particles are entangled, causing temporal correlations of macroscopic quantities of the system. We propose a simple model for these observations.
\end{abstract}

\keywords{Granular avalanche, unjamming, jamming, Lyapunov vector}

-\pacs{83.80.Fg, 45.70.-n, 81.05.Rm, 61.43.-j}
\maketitle

Jamming transition in amorphous materials has become an active research field recently \cite{cates1998jamming, liu1998nonlinear,trappe2001jamming, o2002random,o2003jamming,silbert2006structural, majmudar2007jamming,olsson2007critical,zhang2009thermal,candelier2009creep, chen2010low, zhang2010statistical, xu2010anharmonic, liu2010jamming,van2010jamming, manning2011vibrational, otsuki2011critical,schreck2011repulsive, bi2011jamming, ren2013reynolds,banigan2013chaotic}.The inverse process--the unjamming transition, where a system may suddenly lose rigidity and flow like a liquid, is of crucial importance in studying natural disasters such as snow avalanches, landslides and earthquakes. The continuous tilting of a pile of cohesionless grains will eventually create an avalanche \cite{jaeger1989relaxation, daerr1999two,ramos2009avalanche,amon2013granular,zimber2013polydirectional}, which can be viewed as a dual-process of both the unjamming transition, i.e. when the surface-layer particles lose rigidity and start flowing, and the jamming transition, i.e. when particles come to rest at the end. Granular avalanche has been a paradise of important scientific discoveries such as the Coulomb's laws of friction in the $18th$ century\cite{Duran_book} and the discovery and experimental verification of the Self-Organized Criticality several decades ago\cite{bak1987self,jaeger1989relaxation,frette1996avalanche}. It has important implications in geophysics and in agriculture and industry processes\cite{metcalfe1995avalanche}.

Recent work by Banigan et al understands unjamming and jamming transitions from the novel perspective of the dynamical systems theory\cite{banigan2013chaotic}. They have discovered that the unjamming transition of a system is an unstable fixed point with a strong spatial correlation between the magnitude of the first Lyapunov vector $\delta u(t)$ and the velocity magnitude $v(t)$, whereas the jamming transition is a stable fixed point with a weaker correlation between $\delta u(t)$ and $v(t)$. Such a behavior is very intriguing, showing a distinct characteristics of granular materials compared to thermodynamic systems and glassy systems \cite{shinbrot2013granular}. However the mechanism of the spatial correlation $C_{\delta u,v}$ between $\delta u(t)$ and $v(t)$ is still elusive. Besides, once the transitions take place, the time reversible symmetry of the system is broken\cite{pine2005chaos, corte2008random, keim2014mechanical, regev2013onset, slotterback2012onset}; the irreversibility of transitions poses a great challenge to measure $\delta u(t)$ experimentally. Since $\delta u(t)$ characterizes the degree of divergence or convergence of the evolutionary trajectory of each particle under the most effective perturbations in phase space, it carries critical information of the dynamics of the system that can be crucial in understanding the transition between the static and the dynamical states from the new perspective of dynamical system theory \cite{ott2002chaos}. The present work is for the first time to validate Banigan et al's theoretical model in a real laboratory experiment. One important goal of the present work is to clarify the mechanism of such correlations through the nontrivial measurement of $\delta u(t)$ in a real experiment (See the Supplemental Material for detail).

\begin{figure}
\centerline{\includegraphics[width=1.0\linewidth]{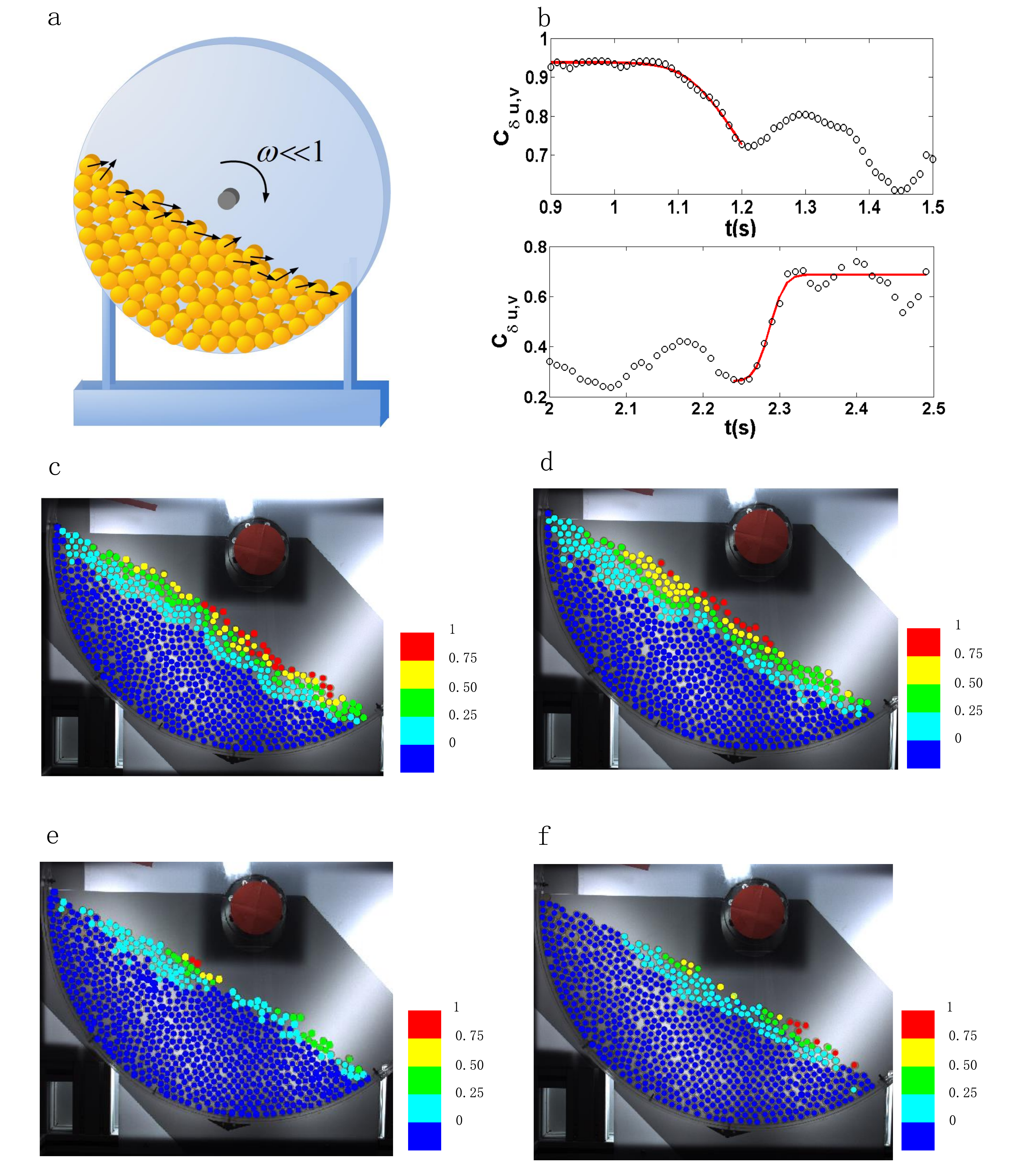}}
\caption{\label{fig:figure1} (a) A schematic of the experimental setup. (b) The correlation $C_{\delta u, v}$ as a function of time in the unjamming (upper panel) and jamming (lower panel) regimes. Red lines are fit using error function to extract the plateau values. The spatial distribution of the modulus of the first Lyapunov vector $\delta u$ (c,e) and the magnitude of the velocity $v$ (d,f) in the unjamming (c-d) and jamming (e-f) regimes. Here the system unjams at $t\approx0.8s$, i.e., the beginning of the avalanche, and jams at $t\approx2.7 s$, i.e., the ending of the avalanche. In panels (c-d) $t=1.3s$ and in panels (e-f) $t=2.4 s$. In the color schemes of panels (c-f), 1 stands for the maximum values of $\delta u$ (in c, e) or $v$ (in d, f).}
\end{figure}

Here we report the direct measurement of $\delta u(t)$ in a model experiment (Fig.~\ref{fig:figure1}(a)), where $\delta u$ and $v$ are strongly correlated in the spatial domain in the unjamming regime, whereas in the jamming regime the correlation is slightly smaller on statistical average. Further analysis shows that the Lyapunov exponents are positive at the unjamming transition and negative at the jamming transition. These results are consistent with the work of Banigan et al\cite{banigan2013chaotic} despite that the two systems are different in terms of the driving: our system is gravity driven with free surfaces and their system is confined and driven by a uniform shear, suggesting that the results could be universal. The setting of our system provides novel insights to understand these results. We further discovered that the dynamics connecting two fixed points of unjamming and jamming is very rich, with the entanglement of the spatiotemporal chaotic fluctuations of velocities of individual particles. As a result, the global Lyapunov vector, its linear growth rate, and the global velocity of the system are strongly correlated temporally.

Our system is essentially a rotating drum as sketched in Fig.~\ref{fig:figure1}(a). It is consisted of a thin cylinder of a diameter of 80 cm, with a rotation axis perpendicular to the direction of the gravity. The rotation speed is typically slow, e.g. 15 minutes per revolution. The cylinder is hollow with a gap of 8 mm between two flat circular plates made of transparent Plexiglas with surfaces coated to reduce the accumulation of electrostatic charges. Inside the cylinder, it is filled with a monolayer of photoelastic disks up to a height slightly over $\frac{2}{3}$  of the radius of the cylinder. These disks, 6.35 mm thick and with a total number of 736, are bidisperse with a large size of 1.4 cm and a small size of 1.2 cm in diameters. The disks are randomly distributed in space with a number ratio of 1:1 to avoid crystallization. During the experimental run, we have observed no particle segregation. The disks are made of the PSM4 materials manufactured by Vishay. The experiment has been repeated for ten times following the identical protocols and the results of independent runs are similar. So here we will present results in two different groups -- results from one randomly selected experimental run to show the details of the dynamics and the statistics of quantitative measures from all ten independent runs to emphasize the common characteristics.

We first analyzed the correlation $C_{\delta u, v}$ between $\delta u(t)$ and $v(t)$ (The details about the measurement of $\delta u(t)$ can be found in the Supplemental Material). The results are shown in Fig.~\ref{fig:figure1}(b-f), where it shows the spatial distributions of $\delta u(t)$ and $v(t)$ at $t=1.3 s$ (in c and d) and at $t=2.4s$ (in e and f) respectively. Here the avalanche starts around $0.8 s$ (the onset of the unjamming transition) and finishes around $2.7 s$ (the onset of the jamming transition). The stable particles are painted in blue, which are excluded in the calculation of $C_{\delta u, v}$. The correlation is defined as $C_{\delta u, v}=\frac{\sum_{i}(\delta u_{i}-\bar{\delta u})(v_i-\bar v)}{\sqrt{\sum_i(\delta u-\bar u)^2} \sqrt{\sum_i(v_i-\bar v)^2}}$, where the summation is over different particle $i$ and $\bar{\delta u}$ (or $\bar {v}$) stands for the average of the quantity.  The results of $C_{\delta u, v}$ in the unjamming (and respectively the jamming) regime are plotted in the upper panel of (b) (and respectively, the lower panel of (b)). Note that the definitions of these two regimes will be discussed in detail later. Also note that $\delta u$ in these two regimes are computed using different ideal trajectories (See Supplemental Material for detail). In the upper panel of (b) , on average $C_{\delta u, v}$ gradually decreases as a function of time; whereas in the lower panel of (b), it first remains flat and then increases rapidly and finally remains flat again. Both curves show fluctuations around 0.1, which are slightly larger in the lower panel than in the upper panel in (b). These curves allow us to extrapolate the values of $C_{\delta u, v}$ at the unjamming and jamming transitions. We note that there is a plateau on the $C_{\delta u, v}$ curve near the unjamming or the jamming transition as seen in Fig.~\ref{fig:figure1}(b). This is a common feature of all runs. We first fit the data points in Fig.~\ref{fig:figure1}(b) using an error function of the form $a*erf(b(x-c))+d$ in the neighbouring regimes of the plateau as drawn using red solid lines in the figure. Here $erf()$ is the error function. We measure the values of the correlation function $C_{\delta u, v}$ around the unjamming and jamming transitions using the plateau values of the fitting. The results are summarized in Table~\ref{table:table1}. From the table, we can see that on statistical average at both the unjamming and jamming transitions the values of $C_{\delta u, v}$ are high. The value of $C_{\delta u, v}$ is slightly larger at the unjamming transition compared to the jamming transition on average. The above results are consistent with Ref.\cite{banigan2013chaotic} despite that the two systems are different in driving-- our system is gravity driven with free surfaces and the system in Ref.\cite{banigan2013chaotic} is confined and driven by uniform shear.

\begin{figure*}
\centerline{\includegraphics[width=0.8\linewidth]{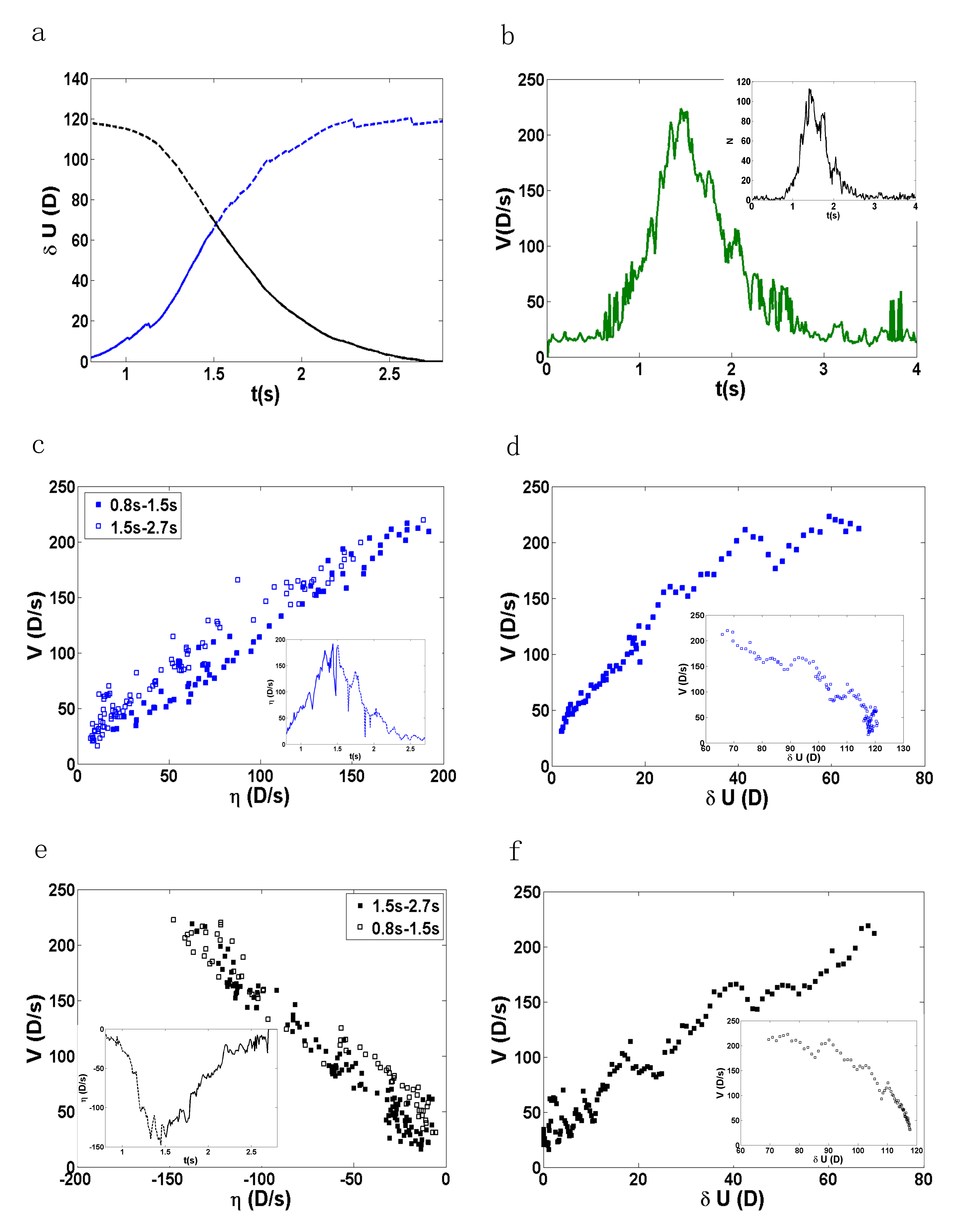}}
\caption{\label{fig:figure2}  (a) The global first Lyapunov vector $\delta U$ versus time. The blue curve is measured with respect to the ideal trajectory of the unjamming transition, whereas the black curve is with respect to the ideal trajectory of the jamming transition. In the blue curve, the solid line is in the unjamming regime and the dashed line extends the curve in the jamming regime. In the black curve, the solid line is in the jamming regime and the dashed line extends the curve in the unjamming regime. (b) The global velocity $V$ versus time. The inset shows the number of avalanche particles $N$ versus time.  (c) $\eta$ versus $V$ inside (solid squares) and outside (open squares) the unjamming regime, where $\eta$ is computed from the blue lines in panel (a) as plotted in the inset. (d) $\delta U$ (the blue lines in (a)) versus $V$ inside (main panel) and outside (inset) the unjamming regime. (e) $\eta$ versus $V$ inside (solid squares) and outside (open squares) the jamming regime, where $\eta$ is computed from the black lines in panel (a) as plotted in the inset. (f) $\delta U$ (the black lines in (a)) versus $V$ inside (main panel) and outside (inset) the jamming regime. Here in the units of $\delta U$, $\eta$, and $V$, $D$ represents the average diameter of the disk.}
\end{figure*}

\begin{table}[h]
\caption{\label{table:table1}
The values of the correlation $C_{\delta u, v}$ at the unjamming and jamming transitions of ten independent experimental runs as specified by the avalanche number $n_0=1,2, ..., 10$. The last two columns are the mean and standard deviation of all the ten runs.
}
{
\scalebox{0.95}{
\scriptsize
\begin{tabular}{|c|c|c|c|c|c|c|c|c|c|c|c|c|c|c|}
\hline
\multicolumn{12}{|c|}  {avalanche number $n_0$} &&\\
\hline
  & $n_0$ & 1 & 2 & 3 & 4 & 5 & 6 & 7 & 8 & 9 & 10 & mean & std \\ \hline
\multirow{2}{*} {$C_{\delta u, v}$} & unjamming &  0.97  & 0.94  & 0.98 & 0.83 & 1.00 & 0.98 & 0.89 & 0.88 & 0.85 & 0.91 & 0.92 & 0.05	\\ 	
               & jamming & 0.90 & 0.69 & 0.90 & 0.88 & 1.00 & 0.77 & 0.65 & 0.97 & 0.83 & 0.88 & 0.84 & 0.11	\\
 												   \hline
\end{tabular}
}
}
\end{table}

Besides the spatial correlation between $\delta u$ and $v$, we also find strong temporal correlations between the global Lyapunov vector of the system $\delta U(t)$, its linear growth rate $\eta(t)$, and the global velocity of the system $V(t)$, as shown in Fig.~\ref{fig:figure2}. Here $\delta U(t)=\sqrt{\sum_i \delta u_i(t)^2}$ describes the deviation from the ideal trajectory of the whole system in phase space at time $t$, $\eta(t)=\frac{d}{dt}\delta U(t)$ and $V(t)=\sqrt{\sum_i v_i(t)^2}$.Note that $\delta u_i(t)$ are measured differently in the unjamming and jamming regimes by referring to different ideal trajectories (See the Supplemental Material for detail) such that there are two different curves of $\delta U(t)$ as shown in Fig.~\ref{fig:figure2}(a). Here the blue solid line represents $\delta U(t)$ in the unjamming regime and the blue dashed line extends the calculation to $t=2.7 s$; similarly, the black solid line represents $\delta U(t)$ in the jamming regime and the black dashed line extends the calculation to $t=0.8 s$. In the above definitions, the summation is over the set of unstable/mobile particles before time $t$. The number of unstable/mobile particles $N(t)$ itself changes dramatically with time, as displayed in the inset of Fig.~\ref{fig:figure2}(b). After $t\approx 0.8 s$, $N(t)$ grows exponentially to its maximum value around $t=1.5 s$ and then starts to decrease exponentially, with the presence of some random fluctuations. Before $t\approx 0.8s$, there are only a few unstable/mobile particles, essentially rattlers, from time to time. We define the unjamming and jamming regimes according to the dynamics of V(t): before (respectively, after) V(t) reaches the peak it is the unjamming (respectively, jamming) regime. By definition, there are two separate curves of $\eta(t)$ in the unjamming and jamming regimes, as shown in the insets of Fig.~\ref{fig:figure2}(c,e) respectively, where $\eta(t)$ and $V(t)$ are strongly correlated in the unjamming and jamming regimes.  In addition, $\delta U(t)$ and $V(t)$ are strongly correlated in both regimes as well , as shown in the main panels of Fig.~\ref{fig:figure2}(d,f), where the insets plot $\delta U$ versus $V$ for $\delta U$ obtained from the extended dashed curves in Fig.~\ref{fig:figure2}(a) just for comparison. Results of the temporal correlations of all ten independent runs have been summarized in Table~\ref{table:table2}. To quantify the linear correlations of these macroscopic variables, we fit the scattered data points using linear fit. The degree of the linear correlations are characterized by the correlation coefficient -- the last number of each cell of the table. A coefficient of $1.0$ or $-1.0$ means a perfect linear correlation between two variables. On average the correlation coefficients are around 0.9, indicating strong linear correlations between $\eta-V$ and $V-\delta U$ in both unjamming and jamming regimes.

\begin{figure}
\centerline{\includegraphics[width=1\linewidth]{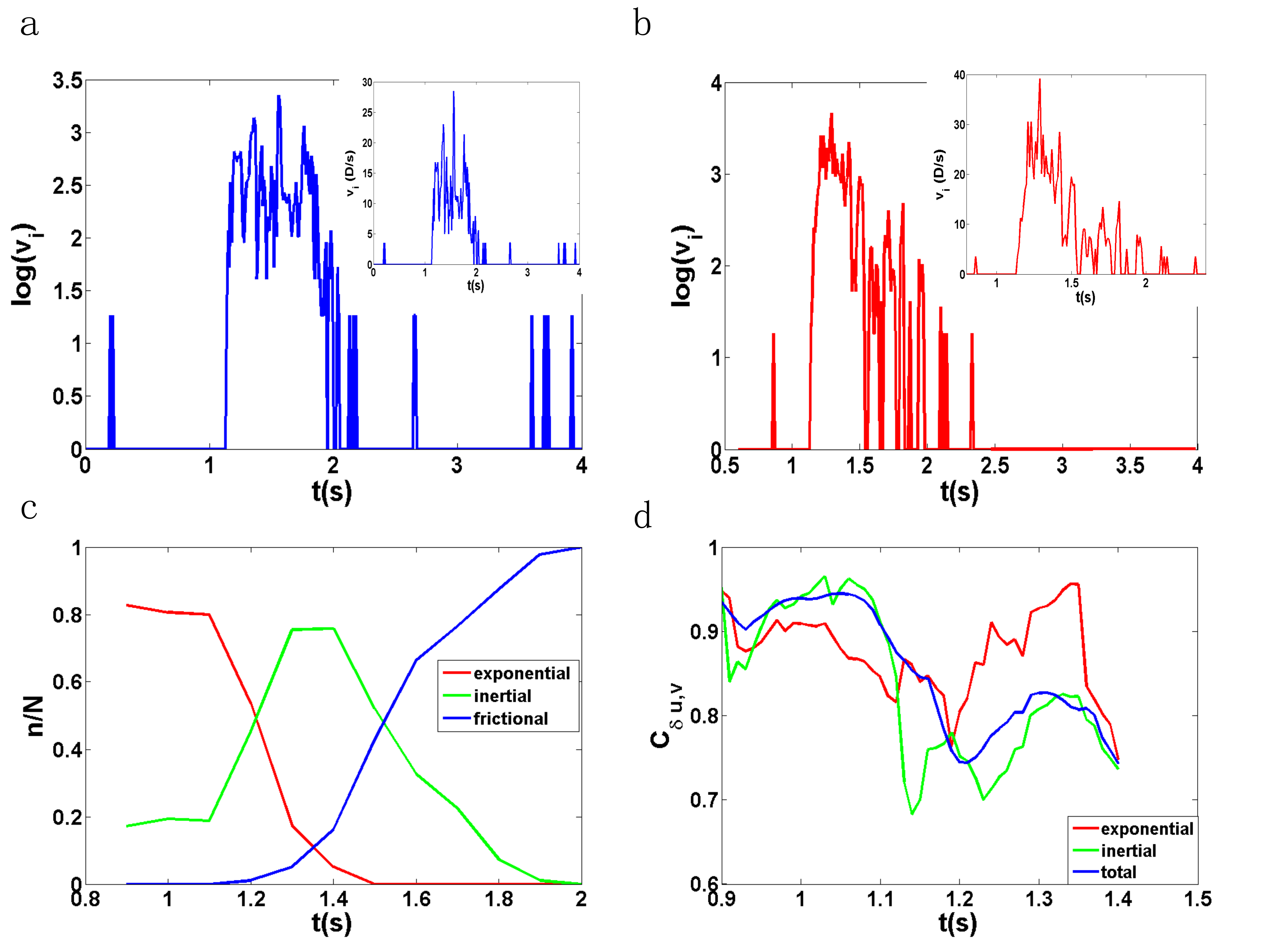}}
\caption{\label{fig:figure3} (a,b)Particle velocity magnitude $v_i$ versus time. Each inset is linear plot. (c) The number ratio $\frac{n}{N}$ in three different velocity regimes versus time. (d) The correlation $C_{\delta u,v}$ of particles in the different regimes of velocity versus time. Here the total means the joint set of particles of the exponential and inertial regimes.}
\end{figure}

In order to understand the above results, a crucial piece of information is the velocity $v_i(t)$ of a single particle $i$, as shown in Fig.~\ref{fig:figure3}(a,b). Besides the two trivial stable regimes where $v_i(t)$ is zero, we distinguish three regimes on this curve as time evolves: (1) a rapid {\it{exponential growth regime}} where the particle loses its stability; (2) an \emph{inertial regime} where the particle continues accelerating till reaching the maximum, accompanied with fluctuations due to random interactions between the particle and others; (3) a {\it{frictional regime}} where the particle loses its kinetic energy and this regime is very bumpy with a lot of fluctuations. In Fig.~\ref{fig:figure3}(c), we plot the number ratio $n/N$ for particles in the above three regimes. From $0.9s$ to $1.5s$ the particles in the exponential and inertial regimes dominate and $n/N$ is non-zero in the exponential regime while the system is rapidly diverging from the unstable fixed point of the unjamming transition with a continuous replenishment of fresh unstable particles. Between $1.5s$ and $2 s$ the inertial and frictional regimes coexist. From $2 s$ to $2.7s$ all the moving particles are in the frictional regime while the whole system is converging towards the stable fixed point of the jamming transition. In Fig.~\ref{fig:figure3}(d), we also plot the spatial correlation $C_{\delta u, v}$ of particles in their individual and joint set of the exponential and inertial regimes.

\begin{table}[h]
\caption{\label{table:table2}
The statistics of the degree of linear temporal correlations between macroscopic variables in ten independent runs. Here `(slp, intc, ccf)' means the slope, intercept, and correlation coefficient of the linear fitting, respectively.
}
{
\scalebox{0.90}{
\scriptsize
\begin{tabular}{ |p{1 cm}|c|c|c|c|c| }
\hline
  avalanche number& & $\eta_{unj}-V$ & $\eta_{jam}-V$ & $\delta U_{unj}-V$ & $\delta U_{jam}-V$ \\ \hline
$n_0=1$ & (slp, intc, ccf)  & 0.95, 29.8, 0.95 & -1.16, 27.3, -0.98 & 2.25, 38.9, 0.97 & 2.21, 38.0, 0.96 \\
 												   \hline
$n_0=2$ & (slp, intc, ccf)  & 1.01, 31.2, 0.89 & -1.33, 21.5, -0.97 & 3.07, 47.7, 0.96 & 2.64, 33.7, 0.97 \\
 												   \hline
$n_0=3$ & (slp, intc, ccf)  & 1.03, 18.6, 0.96 & -1.28, 13.4, -0.95 & 5.82,  16.7,  0.96 & 2.89, 17.0, 0.93 	\\
 												   \hline
$n_0=4$ & (slp, intc, ccf)  & 1.05, 18.5, 0.94 & -1.22, 18.6,  -0.96 & 3.77, 15.5, 0.87 & 2.22, 29.7, 0.95 	\\
 												   \hline
$n_0=5$ & (slp, intc, ccf)  & 1.07, 17.4, 0.93 & -1.21, 14.1, -0.94 & 4.21, 14.3, 0.89 & 1.65, 33.9,  0.88 \\
 												   \hline
$n_0=6$ & (slp, intc, ccf)  & 0.89, 20.9, 0.82 & -1.09, 18.1, -0.74 & 5.10, 6.3, 0.90 & 1.05, 27.0, 0.69	\\
 												   \hline
$n_0=7$ & (slp, intc, ccf)  & 0.98, 23.2, 0.89 & -1.47, 9.0, -0.89 & 2.84, 61.6,  0.76 & 2.51, 26.6, 0.90 \\
 												   \hline
$n_0=8$ & (slp, intc, ccf)  & 0.87, 23.8, 0.86 & -1.14, 16.0, -0.87 & 1.62, 38.2, 0.67 & 1.81, 22.0, 0.90	\\
 												   \hline
$n_0=9$ & (slp, intc, ccf)  & 1.02, 21.2, 0.90 & -1.19, 17.8, -0.93 & 6.80, 9.8, 0.96 & 1.60, 32.5, 0.88 	\\
 												   \hline
$n_0=10$ & (slp, intc, ccf)  & 0.90, 20.0, 0.93 & -1.50, 24.3, -0.91 & 2.14£¬ 18.94, 0.81 & 4.05, -83.6, 0.90 	\\
 												   \hline
mean & (slp, intc, ccf)  & 0.98, 22.5, 0.91 & -1.26, 18.0, -0.91 & 3.77, 26.8, 0.88 & 2.26, 17.7, 0.90 	\\
 												   \hline
\end{tabular}
}
}
\end{table}

To understand the results presented in the above Figs.~(\ref{fig:figure1}-\ref{fig:figure3}), we have proposed a mean-field model, as discussed in detail in the Supplementary Material. First, this model allows us to gain some physical insights to understand the spatial correlation between $\delta u$ and $v$ in the unjamming and the jamming regimes. When a particle becomes unstable, its velocity $v_i(t)$ grows rapidly. As a result, the particle deviates quickly from its original position in real space. In a comparison, at every time instant t the contribution of the ideal trajectory to the measurement of $\delta u_i(t)$ is negligible since the particle would follow the ideal circular motion in a speed much slower than its actual speed. Therefore, ignoring the fluctuations of the moving direction under the mean-field approximation, $\delta u_i(t)$ is approximately an integration of $v_i(t)$, which correlates strongly with $v_i(t)$ in the spatial domain in the unjamming regime. As shown in Fig.~\ref{fig:figure3}(c), in the jamming regime, $v_i(t)'s$ of moving particles are incrementally evolving into the frictional regime, which is much more erratic with large fluctuations, hard to be fully captured by a simple mean field model. This might be the reason that the correlation becomes slightly smaller compared with that in the unjamming regime on statistical average. For the linear correlations between $\delta U$, $\eta$ and $V$ in the unjamming regime, we attribute it largely to the exponential growth of $N(t)$, where more particles become unstable during the cascade of the local unjamming transitions. Similarly, in the jamming regime a constant fraction of moving particles are entering into stable configurations, as caused by the interactions with the stable particles of the neighbouring regions along the pathway of the moving particle--mainly at the downstream of the inclination--to gain stability, creating a cascade of local jamming transitions as the entire system converges rapidly to the stable fixed point. As a result, $N(t)$ decays rapidly in an exponential form. We cannot fully explain the linear correlations between $\delta U$, $\eta$ and $V$ in the jamming regime since the expression of $v_i$ in the jamming regime is obscured by large fluctuations. However, We postulate that the exponential decay of $N(t)$ might be the dominant factor leading to the linear correlations between $\delta U$, $\eta$ and $V$. As a quantitative comparison, we find that the theoretical values of $\delta U$ and $V$ before time $t_1$ agree with the experimental measurement reasonably well though $\eta$ is a little off compared with the real data, which is more sensitive to the parameter values used in the model.As the system evolves from the unstable fixed point at the unjamming transition to the stable fixed point at the jamming transition, there is a spatiotemporal chaotic dynamics, where the global Lyapunov exponent switch signs from positive in the unjamming regime to negative in the jamming regime, consistent with the work of Banigan et al\cite{banigan2013chaotic}.

In conclusion, we have designed a novel experiment which allows us to successfully measure the first Lyapunov vectors $\delta u$ in the unjamming and jamming regimes of granular avalanche processes. This allows us to study the unjamming and jamming transitions from the dynamical systems theory perspective for the first time in the experiment. At the unjamming transition, when particles become unstable, the velocity $v$ of each individual particle grows exponentially fast such that the contributions from the ideal trajectories to the Lyapunov vectors are negligible compared with the real trajectories. Hence $\delta u$ and $v$ are strongly correlated in the spatial domain. When the system rapidly escapes from the unstable fixed point at the unjamming transition to converge to the stable fixed point at the jamming transition, the strong interactions of particles at the surface layers have produced large spatiotemporal chaotic fluctuations, causing a rapid exponential increase and decrease of the number of avalanche particles. As a result, there are strong fluctuations in the velocity dynamics of individual particles in the inertial and frictional regimes compared with the exponential regime, causing a weaker spatial correlation between $\delta u$ and $v$ at the jamming transition. We also have observed strong temporal correlations between the global Lyapunov vector $\delta U$, its linear growth rate $\eta$, and the global velocity $V$, which can be explained reasonably well using a mean-field model. Compared to the recent work of Banigan et al, our results are consistent with their numerical findings, providing supporting experimental evidence for their modeling. The unique setting of our experiment provides new physical insights to explain various correlations in a simple and intuitive way. The drastic difference between two systems in terms of driving suggests the universality of the results. One important question for further study is how to connect the dynamical instability of the system, such as the Lyapunov vector, to the geometrical packing or the force structure of the system. Despite that the grain-scale instabilities and stresses are not correlated\cite{banigan2013chaotic}, success has been achieved to predict, at least statistically, the local rearrangements of particles from the analysis of soft spots\cite{manning2011vibrational,chen2010low}. How to integrate the nonlinear response and the linear response of the system will be something important to investigate in the future.

{\bf Acknowledgments}
J.Z. acknowledges support from the award of the Chinese 1000-Plan (C) fellowship and Shanghai Pujiang Program (13PJ1405300).

\bibliographystyle{apsrev}
\bibliography{dynamical}
\end{document}


\title{Supplementary Material for \lq\lq{}Spatiotemporal Chaotic unjamming and jamming in granular avalanches\rq\rq{}}

\author{Ziwei Wang$^{1}$ and Jie Zhang$^{2,3}$}
\email{jiezhang2012@sjtu.edu.cn}
\affiliation{$^1$Zhiyuan College \& $^2$Institute of Natural Sciences \& $^3$Department of Physics and Astronomy, Shanghai Jiao Tong University, Shanghai 200240, China}

\keywords{Granular avalanche, unjamming, jamming, Lyapunov vector}

-\pacs{83.80.Fg, 45.70.-n, 81.05.Rm, 61.43.-j}

\maketitle

{\bf The measurement of the first Lyapunov vector}

In our experiment, a key physical quantity is the first Lyapunov vector. In general , the computation is rather complex involving the time reversal of the dynamical trajectory of a system in phase space. First, one prepares an initial state and then sets the system to run at time $t=0$ with zero perturbation so that the system evolves freely in phase space. This allows one to identify the \lq\lq{}ideal\rq\rq{} trajectory of the system in phase space, i.e. the trajectory of free evolution of the system under zero perturbation. Second, once the \lq\lq{}ideal\rq\rq{} trajectory is found, one can reverse the time to return the system to the initial state in phase space. Afterwards, one applies a perturbation and records the new evolution trajectory of the system to compute the difference between the new trajectory and the \lq\lq{}ideal\rq\rq{}one. The difference depends on the perturbation. One has to repetedly apply  the time reversal to set the system back to the initial state, and then to apply a different perturbation and let the system evolve in order to find the first Lyapunov vector,i.e. the trajectory where the perturbation grows the fastest. This protocol is much easier to be implemented in computer simulations. However, it is a great challenge in the real granular experiment: (1) one could not have the initial state of a granular system prepared exactly the same; (2) once an unjamming or jamming transition takes place, it is impossible to time-reverse the system to the initial state since the time reversible symmetry is broken \cite{pine2005chaos,corte2008random,keim2014mechanical,regev2013onset,slotterback2012onset}; besides, the \lq\lq{}ideal\rq\rq{} trajectory is often untraceable.

\begin{figure*}
\centerline{\includegraphics[width=0.8\linewidth]{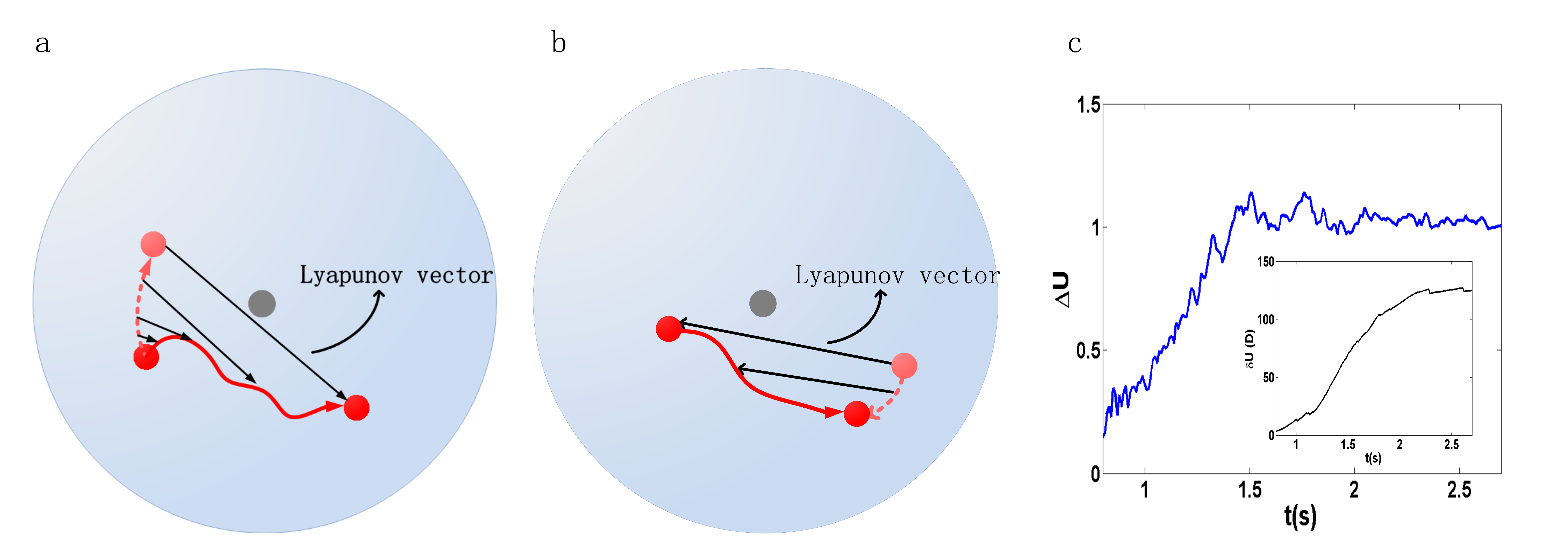}}
\caption{\label{fig:figureS1} Schematic diagrams illustrate how the Lyapunov vectors are defined: (a) during the unjamming transition, i.e. when disks become unstable and start to flow, and (b) during the jamming transition, i.e. when disks stop flowing and come to rest. (c) The modulus of the first global Lyapunov vector $\Delta U(t)$ in the phase space that includes the contribution of the velocity as a function of time $t$ near unjamming. We normalize $\Delta U(t)$ by
$\Delta U(t)=\sqrt{\frac{\sum_{i}\delta u_i(t)^2}{\delta U(t)^2_{max}}+\frac{\sum_{i} v_i(t)^2}{V(t)^2_{max}}}$. In comparison, the inset shows $\delta U(t)=\sqrt{\sum_i\delta u_i(t)^2}$ with no inclusion of the velocity. The units, D, of $\delta U$ is the average diameter of the disk.
}
\end{figure*}

Fortunately, in our system the \lq\lq{}ideal\rq\rq{} trajectory is predictable even after the system undergoes an unjamming or a jamming transition due to external perturbations. We define the unjamming transition of particles as the onset of the avalanche where some particles at the top surface layers lose rigidity and start moving. Similarly, the jamming transition is defined as the flowing particles start coming to rest so that the avalanche comes to the end.  One nice property of our system is that during the unjamming and the jamming transition, the \lq\lq{}ideal\rq\rq{} trajectory of each individual particle is exactly predictable!
If there were no perturbation to trigger the avalanche to take place in advance, each unjammed particle would follow a circular trajectory. Note that there is a subtle difference between the trajectory of a particle in phase space and in real space where the trajectory in phase space includes the velocity contribution. In Fig.~\ref{fig:figureS1}(a-b), we only draw a schematic diagram of a particle\rq{}s trajectories in real space; the velocity of either the \lq\lq{}ideal\rq\rq{} or the actual trajectory can be measured in a straightforward way.

In order to make progress, we have made two assumptions. The first assumption is with regard to the perturbations. External perturbations are unavoidable in a real experiment. Let\rq{}s suppose that after time $t_{0}$, the existence of some perturbations makes the real trajectory diverge from the \lq\lq{}ideal\rq\rq{} one, where the system would reach the maximum angle of repose under zero perturbations. We assume that the existence of perturbations makes the avalanche happen in advance, which is consistent with the experimental observation that the critical angle of repose has a wide range of about 10 degrees\cite{amon2013granular}. Since the duration of the avalanche is short, typically 1 to 2 seconds,  and the rotation speed is slow, i.e. at 15 min per revolution, we believe that the avalanche will finish before the system could have reached the maximum angle of the repose under zero perturbation when it had followed the ideal trajectory.

The second assumption is with regard to the first Lyapunov vector. In our experiment,  it is observed that the trajectories of the avalanche particles have been extremely deviated from the ideal ones. Hence we assume that under perturbations, the difference between the actual trajectory and the \lq\lq{}ideal\rq\rq{} one corresponds to exactly the first Lyapunov vector, which grows the fastest by natural selection.

In contrast to the unjamming transition, where small perturbations grow exponentially, in the jamming transition, small perturbations decrease exponentially to drive the system to a stable fixed point. Therefore, if we apply a time reversal after the avalanche, each particle in the system would essentially follow a circular trajectory in the counter clockwise direction. This can be verified by reversing the rotation of the system counter-clockwise after the avalanche, where the time reversible trajectory indeed follows nicely the circular trajectories. Based on these observations, we believe that the irregular trajectories of moving particles near the end of the avalanche can be treated as the deviation from the \lq\lq{}ideal\rq\rq{} circular trajectories due to perturbations. Here we define the first Lyapunov vector near the jamming transition as the deviation of the real trajectory from the circular ideal one. The above first assumption is also applicable following a similar argument.

{\bf A comment on the Lyapunov vector in phase space}

In the figure 1 of the main text, we have shown that there is a strong spatial correlation between $\delta u$ and $v$ at the unjamming transition, which suggests that a particle will gain more kinetic energy if the particle is more unstable. At the jamming transition, the correlation becomes weaker. In the above measurement of $\delta u$, the deviation between the actual trajectory and the ideal trajectory is computed in real space not in phase space, i.e. no contribution from the velocity part is considered. It is intuitive that considering the contribution of the velocity will only increase the spatial correlation between the two quantities as shown in  the figure 1 of the main text. We also note that adding contributions from the velocity part will make no qualitative difference in the evolution of the global Lyapunov vector $\delta U(t)$ of the system as shown in Fig.~\ref{fig:figureS1}(c).

{\bf The global Lyapunov exponents}

\begin{figure}
\centerline{\includegraphics[width=1.0\linewidth]{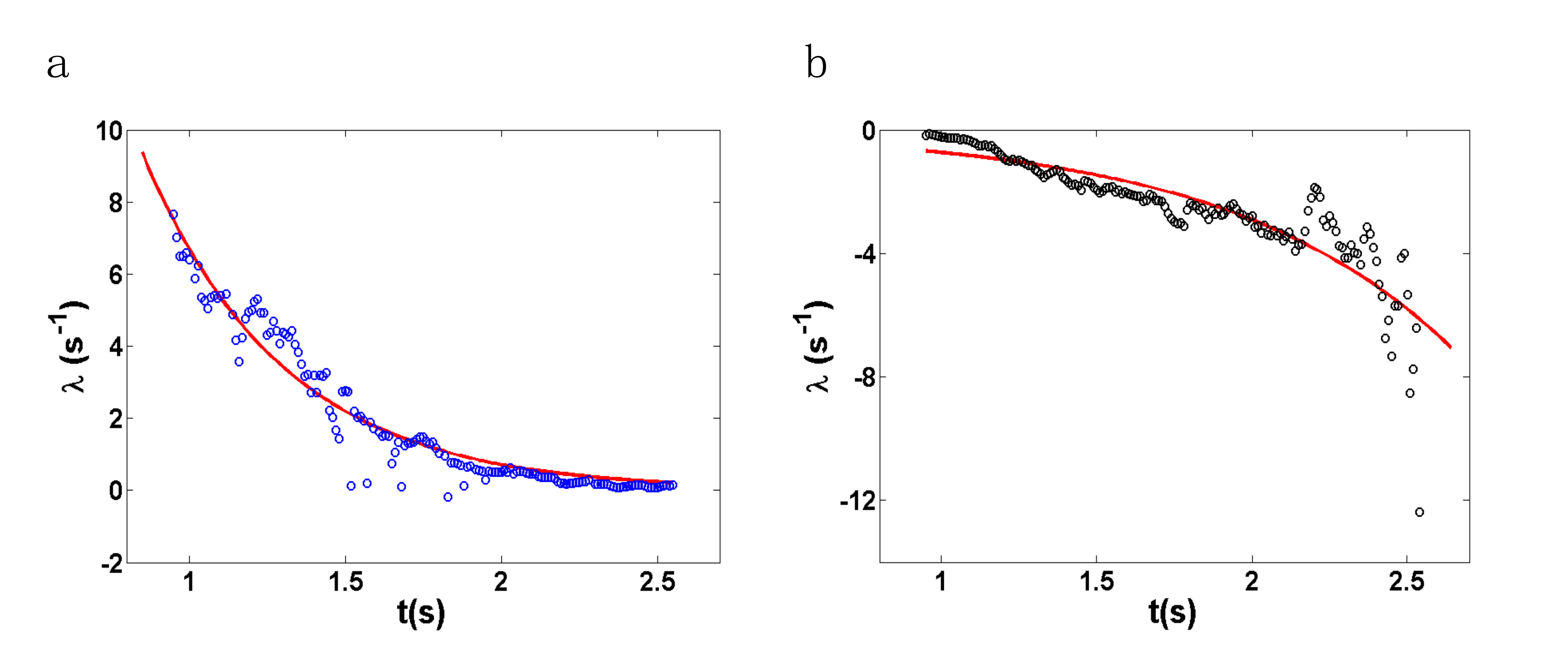}}
\caption{\label{fig:figureS2} The global Lyapunov exponent $\lambda$ in (a) the unjamming regime and in (b) the jamming regime. In each panel, the dashed line extends the calculation of $\lambda$ into the whole range of the avalanche process. Red solid lines are fit using the function  $\lambda(t)=\alpha e^{\beta t}$.}
\end{figure}

To characterize the rate of divergence (respectively, convergence) of the system from a fixed point near the unjamming (respectively, jamming) transition, we define the global Lyapunov exponent as $\lambda=\frac{1}{\Delta t}ln(\frac{\delta U(t)}{\delta U(t-\Delta t)})$, with $\Delta t=0.01 s$ as shown in Fig.~\ref{fig:figureS2}. Here $\lambda$ measures the average exponential growth rate of the global Lyapunov vector $\delta U$ (defined in the main text) over a short time $\Delta t=0.01 s$ which is chosen to be the inverse of the frame rate of the imaging. As shown in the figure, except for a few sharp spikes, most likely due to noises, $\lambda$ is positive in the unjamming regime and becomes negative in the jamming regime.

\begin{table}[h]
\caption{\label{table:table1}
Various fitting coefficients related to the Lyapunov exponent $\lambda$.
}
{
\scalebox{1.0}{
\scriptsize
\begin{tabular}{ |c|c|c|c|c| }
\hline
  avalanche number& $\lambda_{unj}$ & $\lambda_{jam}$ & ($\alpha, \beta$) & ($\alpha, \beta$) \\ \hline
$n_0=1$ &  12.1 & -21.0  & 16.2, -1.9  & -0.0046, 3.4 \\
 												   \hline
$n_0=2$ &  9.3 & -7.0 & 62.6, -2.2  & -0.18, 1.3  \\
 												   \hline
$n_0=3$ &  11.1 & -8.0  & 66.9, -2.3  & -0.29, 1.3 	\\
 												   \hline
$n_0=4$ &  6.7 & -6.5  & 13.7, -1.5 & -0.30, 1.3  	\\
 												   \hline
$n_0=5$ &  6.2 & -8.3  & 16.3, -1.2  & -0.025, 2.3  \\
 												   \hline
$n_0=6$ &  5.5 & -10.2  & 400.0, -1.6  & -0.00074, 2.1 	\\
 												   \hline
$n_0=7$ &  11.2 & -11.5  & 150.4, -4.7  & -0.10, 2.8  \\
 												   \hline
$n_0=8$ &  15.4 & -6.3  & 113.3, -4.4  & -0.30, 1.3 	\\
 												   \hline
$n_0=9$ &  9.3 & -4.7  & 43.0, -2.3  & -0.38, 1.1  	\\
 												   \hline
$n_0=10$ &  8.0 & -3.84 & 871.4, -2.6 & -0.0013, 2.4 \\
 												   \hline
mean &  9.5 & -8.8 & 175.4, -2.5 & -0.16, 2.0	\\
 												   \hline
\end{tabular}
}
}
\end{table}
To make a comparison among all runs, we fit the curves in Fig.~\ref{fig:figureS2}(a-b) with the function of $\lambda(t)=\alpha e^{\beta t}$.Table~\ref{table:table1} summarizes the fitting coefficients $\alpha, \beta$ of different runs of the unjamming and jamming regimes. Note that in each run, the starting point of an avalanche is not zero because experimentally in order to record the whole sequence of the avalanche process, we often start with an arbitrary time shift before the actual starting of the avalanche. This is reflected in the wide range of fluctuation of the fitting parameter $\alpha$. We may shift the time $t$ to $t'$ such that the avalanche starts at $t'=0$ and the fitting function becomes  $\lambda(t')=\lambda_{unj}e^{\beta t'}$ for panel (a); we may also shift the time $t$ to $t''$ such that the avalanche ends at $t''=0$ and $\lambda(t'')=\lambda_{jam}e^{\beta t''}$ for panel (b). From Table~\ref{table:table1}, for all runs the values of $\lambda_{unj}$ are positive and $\lambda_{jam}$ are negative, which is robust. For different runs, $\lambda_{unj}$ or $\lambda_{jam}$ are within the same order of magnitude despite the fluctuations. Compared to the mean values, individual values could reach twice larger or smaller.

{\bf The mean-field model}

In order to explain the spatial correlations between $\delta u$ and $v$ and the temporal correlations between $\delta U$, $\eta$ and $V$ in the unjamming regime, we propose a mean-field model based on the dynamics of $v(t)$ and $N(t)$.

{\bf (a) The spatial correlation between $\delta u_i(t)$ and $v_i(t)$}

As discussed in detail in the main text, the velocity $v_i(t)$ of a given particle $i$ in the avalanche can be divided into three regimes -- the exponential regime, the inertial regime, and the frictional regime. If the particle is in the early time range of the unjamming regime, its velocity $v_i(t)$ can be simply modeled as two parts -- an exponential part followed by an inertial part so that
$$v_i(t)=
\begin{cases}
v_0e^{at}& \text{$0<t<t_0$}\\
bt& \text{$t_0<t<t_1$}
\end{cases}$$
where $t_0$ (respectively, $t_1$) is the starting point of the inertial (respectively, the frictional) regime and $v_0=2.5 D/s$, with $D$ the average diameter of the disk, is the minimum resolvable velocity in the experiment. Note that in the figure 4(c) of the main text, for time $t<1.4 s$ the fraction of particles in the frictional regime is less than ten percent and thus the exclusion of the frictional regime in the above model is statistically valid. Since the ideal trajectory contributes negligibly compared with the real one to the first Lyapunov vector $\delta u_i$, we can simply approximate $\delta u_i$ using its actual displacement in the real space, which allows us to obtain $\delta u_i$ by integration:
$$\delta u_i(T)=\int_0^T v_i(t)dt$$
$$=\begin{cases}
v_0(e^{aT}-1)\frac{1}{a}\approx v_0e^{aT}\frac{1}{a}=\frac{1}{a}v_i(T). & \text{$0<T<t_0$}\\
v_0(e^{at_0}-1)\frac{1}{a}+\frac{1}{2}b(T^2-t_0^2)\approx \frac{1}{2}bT^2=\frac{1}{2}\frac{v_i^2(T)}{b}. & \text{$t_0<T<t_1$}
\end{cases}$$
So $\delta u_i(T)$ and $v_i(T)$ are spatially correlated in the leading order whether $v_i(t)$ is in the exponential regime or in the inertial regime.

{\bf (b) The temporal correlations between $\delta U$, $\eta$, and $V$}

In the unjamming regime, individual moving particle can trigger particles of the neighbouring regions along its pathway -- mainly at the upstream of the inclination -- to lose stability, causing a cascade of local unjamming transitions as the whole system is rapidly diverging from the unstable fixed point. As a result, $N(t)$ increases substantially, and assuming a constant growth rate $\xi$ we have
$$N(t)=e^{\xi t}$$
which is consistent with the inset of the figure 2(b) of the main text, with $\xi=7 s^{-1}$.

To find $\delta U(t)$, we compute the convolution of $N(t)'$ and $\delta u_i(t)$ by assuming each mobile particle behaves similarly, so that
$$\delta U(T)=\int_0^T N(t)'\delta u_i(T-t)dt=\int_0^T \xi e^{\xi t}\delta u_i(T-t)dt$$
$$=
\begin{cases}
\frac{\xi v_0}{a(\xi-a)}(e^{\xi T}-e^{aT}), & \text{$0<T<t_0$}\\
C_1e^{\xi T}+C_2T^2+C_3T+C_4, & \text{$t_0<T<t_1$}
\end{cases}$$
with
$$C_1=(\frac{b}{2}t_0^2+\frac{b}{\xi}t_0+\frac{b}{\xi^2})e^{-\xi t_0}+\frac{\xi v_0}{a(\xi-a)}[1-e^{(a-\xi)t_0}]$$ $$C_2=-\frac{b}{2}, C_3=-\frac{b}{\xi}, C_4=-\frac{b}{\xi^2}$$

Taking derivative of $\delta U(T)$, we have $\eta(T)$ as follows
$$\eta(T)=\delta U(T)'=
\begin{cases}
\frac{\xi v_0}{a(\xi-a)}(\xi e^{\xi T}-a e^{aT}), & \text{$0<T<t_0$}\\
D_1e^{\xi T}+D_2T+D_3, & \text{$t_0<T<t_1$}
\end{cases}$$
with
$$D_1=(\frac{b}{2}\xi t_0^2+bt_0+\frac{b}{\xi})e^{-\xi t_0}+\frac{\xi^2 v_0}{a(\xi-a)}[1-e^{(a-\xi)t_0}]$$
$$D_2=-b, D_3=-\frac{b}{\xi}$$
Similarly, $V(T)$ of the system is the convolution of $N(t)'$ and $v_i(t)$:
$$V(T)=\int_0^T N_i(t)'v_i(T-t)dt=\int_0^T \xi e^{\xi t}v_i(T-t)dt$$
$$=
\begin{cases}
\frac{\xi v_0}{\xi-a}(e^{\xi T}-e^{aT}), & \text{$0<T<t_0$}\\
E_1e^{\xi T}+E_2T+E_3, & \text{$t_0<T<t_1$}
\end{cases}$$
with
$$E_1=(bt_0+\frac{b}{\xi})e^{-\xi t_0}+\frac{\xi v_0}{\xi-a}(1-e^{-\xi t_0}), E_2=-b, E_3=-\frac{b}{\xi}.$$

\begin{figure}
\centerline{\includegraphics[width=1.0\linewidth]{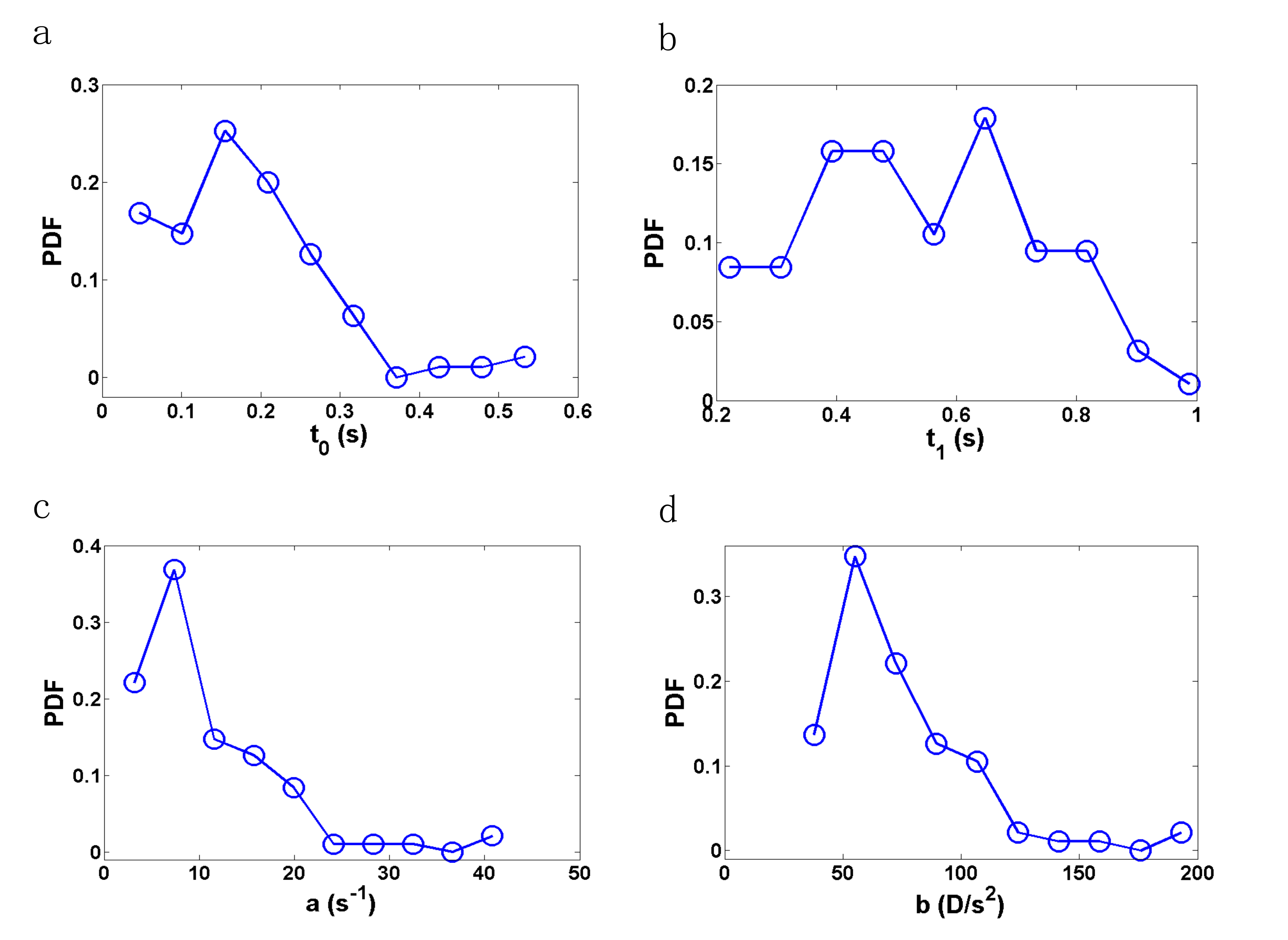}}
\caption{\label{fig:figureS3} The probability distribution functions of various parameters defined in the theoretical model, including (a) $t_0$, (b) $t_1$, (c) $a$, and (d) $b$, respectively. In panel (d), in the units of $b$, $D$ is the average diameter of the disk. }
\end{figure}

\begin{figure}
\centerline{\includegraphics[width=1.0\linewidth]{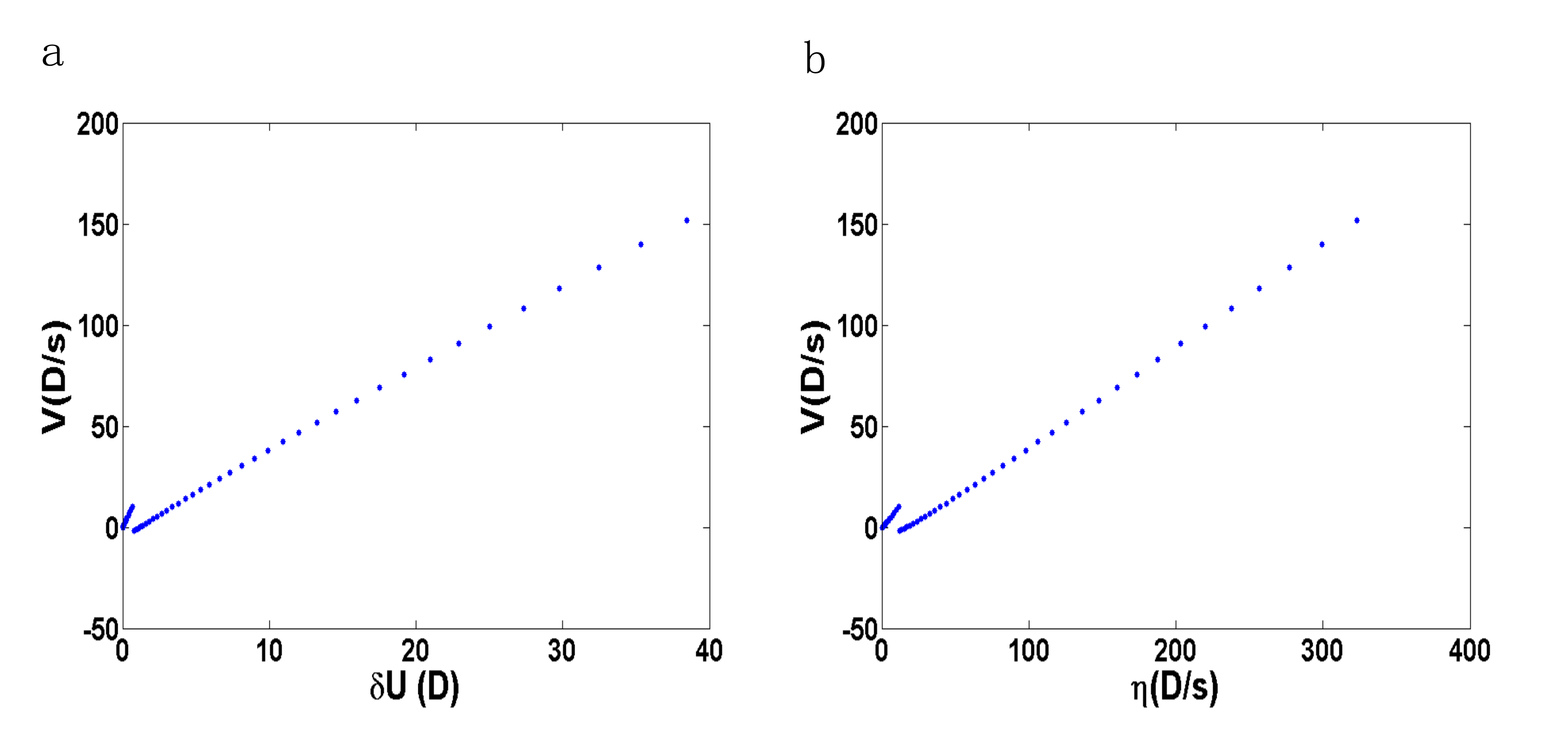}}
\caption{\label{fig:figureS4} Results calculated using our theoretical model: (a) $\delta U$ versus $V$ inside the unjamming regime (b) $\eta$ versus $V$ inside the unjamming regime. Here in the units of $\delta U$, $\eta$, and $V$, $D$ is the average diameter of the disk.}
\end{figure}

\begin{figure*}
\centerline{\includegraphics[width=0.8\linewidth]{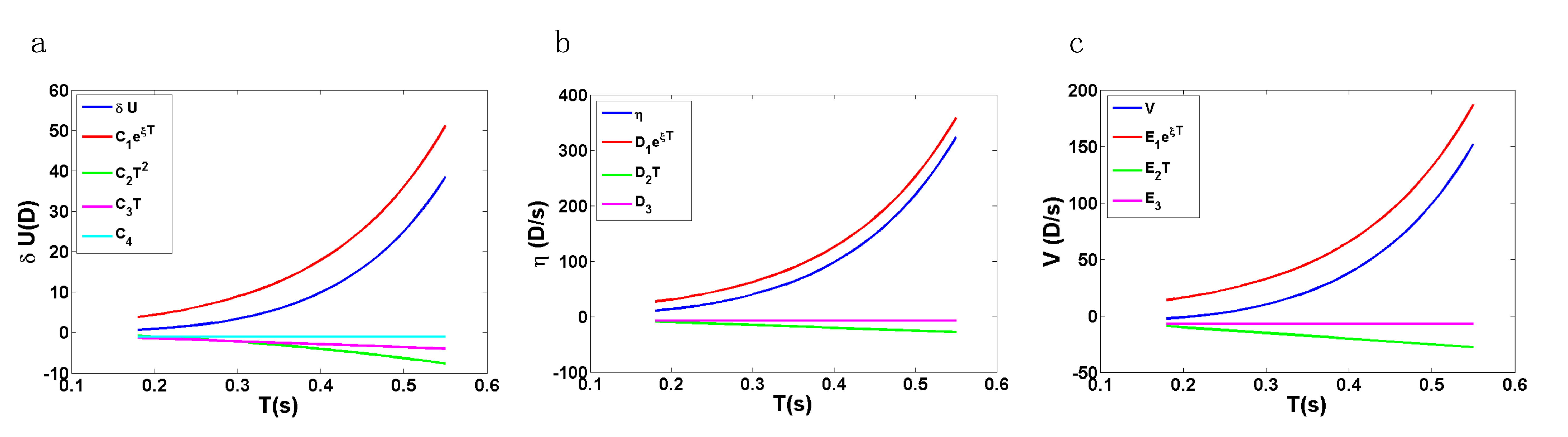}}
\caption{\label{fig:figureS5} (a) $\delta U$ versus time $t$ (blue curve) along with the time curves of the four separate terms in the expression of $\delta U$. (b) $\eta$ versus time $t$ (blue curve) along with the time curves of the three separate terms in the expression of $\eta$. (c) $V$ versus time $t$ (blue curve) along with the time curves of the three separate terms in the expression of $V$. All results here are calculated using the theoretical model. Here in the units of $\delta U$, $\eta$, and $V$, $D$ is the average diameter of the disk.}
\end{figure*}

\begin{table}[h]
\caption{\label{table:table2}
Various parameters used in the mean field model.
}
{
\scalebox{1.0}{
\scriptsize
\begin{tabular}{ |c|c|c|c|c|c| }
\hline
  avalanche number& $t_0$ & $t_1$ & a & b & $\xi$ \\ \hline
$n_0=1$ &  0.059 & 0.53 & 23.4 & 20.4 & 4.0 \\
 												   \hline
$n_0=2$ &  0.17 & 0.54 & 10.6 & 73.5 & 7.0 \\
 												   \hline
$n_0=3$ &  0.079 & 0.21 & 16.7 & 70.7 & 7.7	\\
 												   \hline
$n_0=4$ &  0.053 & 0.28 & 15.0 & 78.9 & 5.9	\\
 												   \hline
$n_0=5$ &  0.036 & 0.12 & 56.2 & 193.7 & 4.1 \\
 												   \hline
$n_0=6$ &  0.058 & 0.22 & 15.3 & 68.7 & 4.3	\\
 												   \hline
$n_0=7$ &  0.056 & 0.20 & 16.1 & 84.5 & 12.6 \\
 												   \hline
$n_0=8$ &  0.043 & 0.16 & 24.1 & 107.8 & 5.2	\\
 												   \hline
$n_0=9$ &  0.062 & 0.29 & 15.6 & 86.3 & 11.1 	\\
 												   \hline
$n_0=10$ &  0.030 & 0.11 & 49.9 & 246.7 & 2.9	\\
 												   \hline
mean &  0.065 & 0.22 & 20.2 & 85.9 & 5.4	\\
 												   \hline
\end{tabular}
}
}
\end{table}

In order to test the validity of our theoretical model, we calculate $\delta U$, $V$ and $\eta$ using the expressions introduced in the main text with the typical parameters determined from the experiment: $t_0=0.18s, t_1=0.55s, a=10.6 s^{-1}, b=73.5 D/s^2, \xi=7 s^{-1}$. Except $\xi$ which is estimated from the figure 2(b) of the main text, the values of the other parameter are obtained from the statistical averages according to their probability distribution functions shown in Fig.~\ref{fig:figureS3}. The results are plotted in Fig.~\ref{fig:figureS4}, showing strong correlations between $\delta U$, $V$, and $\eta$, which is less obvious from the direct examination of their expressions. So the correlation between these three macroscopic variables can be well captured by our model. There is a small discontinuity near the beginning of the curve in Fig.~\ref{fig:figureS4} due to the simplification in our model where an effective coefficient $b$ is applied to approximate the linear relation of $v_i(t)$ after $t_0$, causing a discontinuity in $v_i$ and subsequently all the derived macroscopic variables at $t_0$. In table~\ref{table:table2}, we have listed all the five parameters used in the above mean field model for different runs. The values of each set of parameters vary from run to run with fluctuations up to a few times larger or smaller compared to the mean value. Note that the formulation of this model is based on two important pieces of information from experimental measurements: (1) the velocity of individual particles can be simply divided into three regimes -- the exponential, inertial and frictional regimes, which allows us to distinguish the jamming and unjamming regimes; (2) the number of the moving particles $N(t)$ increases (respectively, decays) exponentially in the unjamming (respectively, jamming) regime. Using the typical parameter values, in the range of $t_0< T< t_1$, the exponential terms $C_1 e^{\xi T}$, $D_1 e^{\xi T}$, and $E_1 e^{\xi T}$ make the dominant contributions in the expressions of $\delta U$, $\eta$, and $V$,  as plotted in Fig.~\ref{fig:figureS5}. Hence it is the exponential form of $N(t)$ that contributes the most to the linear correlations seen above.